\documentstyle[11pt]{article}  
  
\begin{document}

\title{Extinctions in the random replicator model}

\author{Viviane M. de Oliveira and 
J. F. Fontanari \\
 Instituto de F\'{\i}sica de S\~ao Carlos \\
 Universidade de S\~ao Paulo \\
 Caixa Postal 369 \\
 13560-970 S\~ao Carlos SP, Brazil 
}

\date{}

\maketitle

\begin{abstract}
The statistical properties of an ecosystem composed of
species interacting via pairwise, random interactions and
deterministic, concentration limiting self-interactions are
studied analytically with tools of equilibrium statistical
mechanics of disordered systems. Emphasis is given to 
the effects of  externally induced extinction of a fixed fraction 
of species at the outset of  the evolutionary process. The manner 
the ecosystem copes with the initial extinction event depends on the
degree of competition among the species as well as on the strength of
that event.  For instance, in the regime of high competition
the ecosystem diversity, given by the fraction of surviving species, is
practically insensitive to the strength of the initial extinction provided
it is not too large, while in the less competitive regime  the
diversity decreases linearly with the size of the event. 
In the case of  large extinction events we find that no further biotic
extinctions  take place and, furthermore,  that rare 
species become very unlikely  to be found in the ecosystem at equilibrium. 
In addition, we show that
the reciprocal of the Edwards-Anderson order parameter yields 
a good measure of the diversity of the model ecosystem.
\end{abstract}

\bigskip

PACS: 87.10+e, 87.90.+y, 89.90+n

\newpage

\small 

\section{Introduction}\label{sec:level1}

Extinction seems to be the final outcome of the evolution of species. 
In fact, as species survive about 10 million years  in the average, nearly all 
species that have ever existed are extinct and only a very small
fraction of them have left their impressions in the fossil record \cite{Raup,Newman1}. 
The causes of the mass extinction  events is currently a matter of dispute as
there are two main types of explanations \cite{Newman2}. 
The more
traditional one asserts that extinction is caused by external stresses
as, for instance, major climate changes and asteroids impacts. This point of view is 
supported by some evidences such as 
the unusual quantity of iridium and other noble metals in the rocks
that marked the boundary between the Cretaceous and Tertiary periods,
when the era of the dinosaurs was replaced by the era of the mammals \cite{Alvarez}. 
Since iridium is more common in asteroids than in the Earth's crust, this finding
can be viewed as evidence for an asteroid impact. The alternative explanation
asserts that extinctions are caused by the interactions between the
species in the ecosystems. In particular, Paine \cite{Paine} has shown that species
richness sometimes can be increased by the predator-mediated coexistence, and
the removal of predators can lead to additional species extinctions. Some recent
studies indicate that food webs with
many species or high connectivity are more likely to lose species as
a consequence of the extinction of a single species when compared with 
more simple food webs \cite{Pimm1,Pimm2}. Although this kind of argument seems
well suited to explain the so-called background extinctions, it certainly 
needs some new ingredients to explain mass extinctions as well. In fact,
the missing ingredient seems to be the self-organized criticality 
concept \cite{SOC}, which in this context is best illustrated by the
popular Bak-Sneppen model \cite{Bak}. According to this model,
the fitness of each species is affected by the other
species to which it is coupled in the ecosystem so that
large events in the evolutionary history may be thought of as large
coevolutionary avalanches caused by the intrinsic dynamics of the model.
In this  model, the distribution of the extinction
sizes follows a power law, which is a valid candidate for fitting to 
the experimental data.

Although we recognize that evolution and hence 
extinction are, as pictured
by the models mentioned above, essentially dynamical phenomena, in this 
work we study
these phenomena within the equilibrium statistical mechanics framework of 
the random replicator model for species coevolution 
\cite{Opper,Biscari,novo,Oliveira}. 
Deterministic replicator models are commonly used to describe
the  evolution of self-reproducing entities in a variety of fields such as game
theory, prebiotic evolution and sociobiology \cite{Schuster,hs-book}.
The random  replicator model introduced by Diederich and Opper \cite{Opper}
attempts to model the uncertainties and the overwhelming complexity of
the interspecies interactions in biological ecosystems by assuming that
those  interactions are random. However, it is
also assumed that the dynamics is such that a fitness functional 
(Lyapunov function) is maximized so that the only stationary states  
are fixed points. In fact, the existence of
such a functional leads to a replicator equation with symmetric 
interspecies interactions \cite{hs-book} which is
a severe assumption  from the biological standpoint. It allows, however, 
full use of tools of the equilibrium statistical mechanics 
to study 
analytically the  average properties of the equilibrium states of
this kind of disordered ecosystems. 

An interesting result of the
random replicator model  is that in the equilibrium state
a fraction of the species is extinct \cite{Opper}.
The mechanism of extinction
is clearly outcompetition and, in the absence of any cooperation pressure,
only the pair of species with the largest reinforcing interactions
will thrive. 
In this contribution we study the effects of random elimination
of a fixed fraction of the species at the outset of the evolutionary
process, giving emphasis to the  
distribution of the remaining species concentrations in the equilibrium state.
There are several interesting issues that can be addressed in this
framework. For instance, what is the equilibrium situation when
the fraction of species eliminated at the beginning is
already larger than the fraction that would be extinct naturally
due to outcompetition? Furthermore, how does the ecosystem cope with
large initial extinction events? In this paper we give 
clear-cut analytic answers to these questions which are
partly corroborated by numerical simulations of the model ecosystem.

The remainder of the paper is organized as follows.
In Sec.\ \ref{sec:level2} we introduce the model and discuss
the ecological interpretation of the control parameters. The
equilibrium properties of the model are derived within the replica-symmetric
framework in Sec.\ \ref{sec:level3} and the use
of the reciprocal of the Edwards-Anderson order parameter as a measure
for the diversity of the ecosystem is suggested. In Sec.\ \ref{sec:level4}
we calculate the distribution of probability of the concentrations of
a given species, allowing thus the explicit calculation of the
ecosystem diversity as the fraction of surviving species at equilibrium. 
Finally, in Sec.\ \ref{sec:level5} we present some concluding remarks.

%
\section{Model}\label{sec:level2}
%

We consider an infinite population (ecosystem) composed of 
individuals belonging to $N$ different species whose fitness 
${\mathcal F}_i ~(i=1, \ldots,N )$ are
the derivatives ${\mathcal F}_i = \partial {\mathcal F}/\partial x_i$
of the fitness functional ${\mathcal F}$ defined as
\begin{equation}\label{H_p}
- {\mathcal F} =  {\mathcal{H}} \left ( {\mathbf x} \right ) = 
u ~\sum_i b_i x_i^2 ~+
 \sum_{i <j}
J_{ij} \, b_i x_{i} b_j x_j 
\end{equation}
where $x_i \in [0, \infty) $  is the fraction of species $i$ and 
$b_{i}$ is a quenched random variable that takes the values $0$ and $1$
with probabilities $a$ and $1-a$, respectively. Hence 
$Na$ randomly chosen species are eliminated at the outset in the average
and so henceforth we will refer to $a \in [0,1]$ as the dilution parameter.
An effective competition among the species is enforced by 
requiring that  the concentrations of the surviving species
satisfy the constraint
\begin{equation}\label{constraint}
\sum_{i=1}^N b_i x_i = Q_0 N,
\end{equation}
\noindent
where $Q_0$ is an arbitrary positive constant which gives the scale of
the concentrations $x_i$.
The coupling strengths $J_{ij}$ between species $i$ and $j$
are statistically independent 
quenched random variables with a Gaussian distribution
\begin{equation}\label{prob}
{\cal{P}} \left ( J_{ij} \right ) =
\sqrt{\frac{N}{2\pi }} \exp \left [ 
-\frac{ \left( J_{ij} \right)^2 N}{2} 
\right ]
\end{equation}
so that $J_{ij} < 0$ corresponds to pairs of cooperating species
while $J_{ij} > 0$  to pairs of competing species.
The self-interaction parameter 
$u \geq 0 $ acts as a global cooperation pressure limiting the growth of 
any single species, and it is crucial to guarantee the existence 
of a nontrivial thermodynamic limit, $N \rightarrow \infty$. 
In fact, for large $u$ the minimum of ${\mathcal{H}}$ corresponds
to a homogeneous ecosystem where the surviving species have concentrations
 $x_i/Q_0 = 1/(1-a)~\forall i$. The  positive self-interactions 
means that individuals of a same species compete against themselves,
which is quite reasonable as they certainly  share the same
resources (ecological niche).

The time evolution of the species concentrations is given by  
the replicator equation
\begin{equation}\label{eq:rep}
\frac{dx_i}{dt}=-x_i \left[ \frac{\partial \mathcal{H}(\mathbf{x})}{\partial
x_i} - \frac{1}{N} \sum_k x_k \frac{\partial \mathcal{H}
(\mathbf{x})}{\partial x_k} \right]  \forall i 
\end{equation}
which minimizes $\mathcal{H}(\mathbf{x})$ while keeping the term
$\sum_i b_i x_i$ constant during the evolution. Hence the fixed 
points of this equation are the minima of $\mathcal{H}(\mathbf{x})$ and
in the following we use the replica formalism to study analytically
the statistical  properties of these   minima.

\section{The replica approach}\label{sec:level3}

Following the standard prescription of performing quenched averages
on extensive quantities only \cite{MPV}, we define the average
free-energy density $f$ as
\begin{equation}\label{f0}
- \beta f = \lim_{N \rightarrow \infty} \frac{1}{N}  \left \langle 
\ln Z \right \rangle 
\end{equation}
where 
\begin{equation}\label{Z0}
Z = \int_0^\infty \prod _i dx_i ~ \delta \left ( Q_0 N - \sum_i b_i x_i
\right ) \left ( QN - \sum_i x_i(1-b_i)
\right ) \mbox{e}^{- \beta {\mathcal H} 
\left ( {\mathbf x} \right ) }
\end{equation}
is the partition function and $\beta = 1/T$ is 
the inverse temperature. Taking the limit $T \rightarrow 0$ in 
Eq.\ (\ref{Z0}) ensures that only the states that minimize 
${\mathcal H} \left ( {\mathbf x} \right )$
will contribute to $Z$.
We impose the additional constraint
\begin{equation}
\sum_i x_i (1-b_i)=QN
\end{equation}
to avoid divergences when carrying out the integrals over $x_i$. 
Here $ \langle \ldots \rangle $ stands for the average over
the coupling strengths $J_{ij}$ as well as over the auxiliary variables
$b_i$. As usual, the evaluation of the
quenched average in Eq.\ (\ref{f0}) can be carried out through the
replica method: using the identity
\begin{equation}
\langle \ln Z \rangle = \lim_{n \rightarrow 0} \frac{1}{n} \ln \langle Z^n
\rangle \end{equation}
we first calculate $\langle Z^n \rangle$ for {\it integer}  $n$, i.e., 
$Z^n=\prod_{\rho=1}^{n} Z^\rho$, and then analytically continue to $n=0$. The final
result is
\begin{eqnarray}\label{f_gen}
-\beta f &=& \lim_{n \rightarrow 0} \mbox{extr} \frac{1}{n} \bigg\{
\sum_{\rho}\hat{p}^\rho p^\rho - \frac{\beta u}{2} \sum_\rho p^\rho + \frac{\beta^2}{4}
\sum_\rho (p^\rho)^2 + \sum_{\rho<\delta} \hat{q}^{\rho\delta} q^{\rho\delta} 
+ \sum_{\rho} Q \hat{Q}^\rho \nonumber\\ 
&&+ \frac{\beta^2}{2} \sum_{\rho<\delta}
(q^{\rho\delta})^2+ \sum_{\rho} Q_0 \hat{R}^\rho  + 
\sum_{b=0}^1 P_b \ln G_0(b, \hat{p}^\rho,\hat{q}^{\rho\delta},\hat{R}^\rho,\hat{Q}^\rho)
\bigg\}   
\end{eqnarray}
where $P_0 = a$, $P_1 = 1-a$, and
\begin{eqnarray}\label{G0}
G_0 &=&  \int_{0}^{\infty} \prod_\rho
dx^\rho \exp \bigg\{ -b \sum_\rho \hat{p}^\rho (x^\rho)^2 - b \sum_{\rho<\delta}
\hat{q}^{\rho\delta} x^\rho x^\delta  \nonumber\\
&& - b \sum_\rho\hat{R}^\rho x^\rho - \left ( 1-b \right ) 
\sum_\rho \hat{Q}^\rho x^\rho  \bigg\} .
\end{eqnarray}
We note that while we have calculated the average over the couplings 
$J_{ij}$ explicitly,
we have used the self-averaging property $\frac{1}{N} \sum_i \ln G_0 (b_i) =
\sum_b P_b \ln G_0 (b)$ to eliminate the site dependence of the
$b_i$ variables.
The relevant physical order parameters are
\begin{eqnarray}
q^{\rho\delta} & = & \frac{1}{N}  \sum_i  \langle \langle x_i^\rho x_i^\delta  
\rangle_T \rangle         ~~~~~~~\rho<\delta\\
p^{\rho} &  = & \frac{1}{N} \sum_i \langle \langle \left ( x_i^\rho \right )^2  
\rangle_T \rangle
\end{eqnarray}
which
measure the overlap between a pair of different equilibrium states 
${\bf x}^\rho$ and ${\bf x}^\delta$, and the overlap of an 
equilibrium
state ${\bf x}^\rho$ with itself, respectively. Here, 
$\langle \ldots \rangle_T$ stands for a thermal average taken 
with the Gibbs  probability distribution
\begin{equation}\label{Gibbs}
{\mathcal W} \left ( {\mathbf x } \right ) = \frac{1}{Z} 
~\delta 
\left (Q_0 N - \sum_i x_i \right ) 
~\delta 
\left (Q N - \sum_i (1-b_i) x_i \right )\; \exp \left [ - \beta
{\mathcal H} \left ( {\mathbf x} \right ) \right ] .
\end{equation}

\begin{figure}
\includegraphics{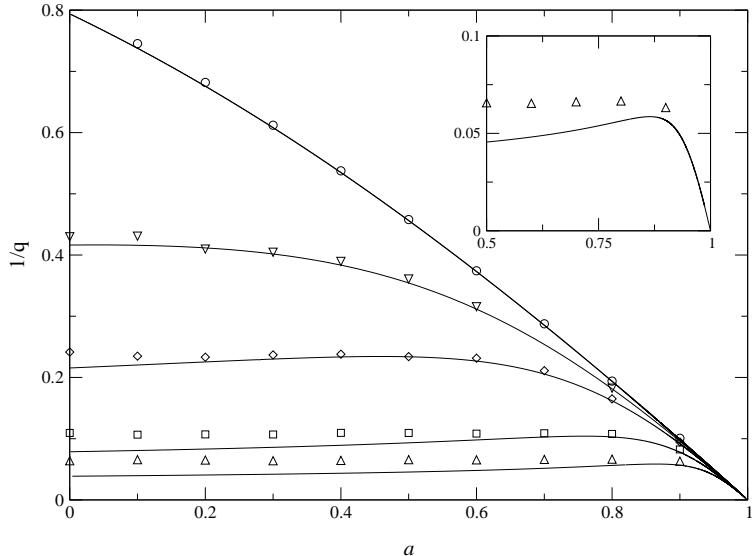} 
\vspace{6.5cm}
\caption{The diversity $1/q$ as a funcion of the dilution parameter
$a$ for  (top to bottom) $ u= 1.3 ~(\bigcirc), 0.8 ~(\bigtriangledown),
0.6 ~(\Diamond), 0.4 ~(\Box)$ and $ 0.3 ~(\bigtriangleup)$. 
The symbols are the results of the numerical solution of the
replicator equation. The inset highlights the region of the diversity
maximum.} 
\label{fig:q}
\end{figure}

\begin{figure}
\includegraphics{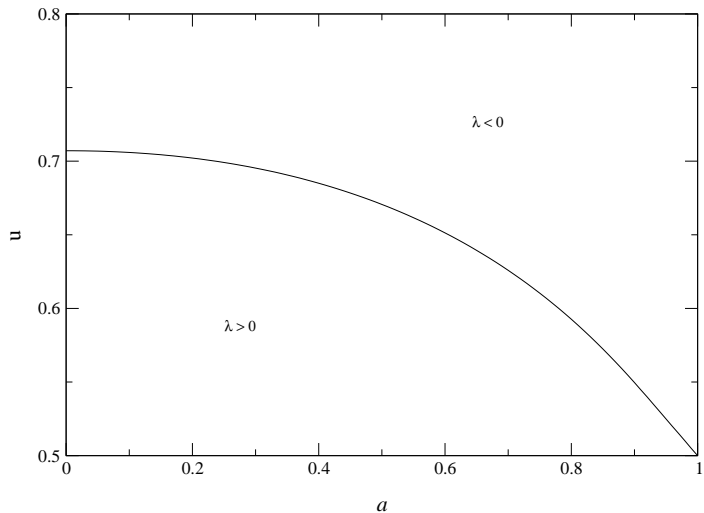} 
\vspace{6.5cm}
\caption{Almeida-Thouless line separating the regions of stability ($\lambda <0$)
and instability ($\lambda > 0$) of the replica-symmetric solution.}
\label{fig:AT}
\end{figure}

To proceed further we assume that the saddle-point parameters are symmetric under
permutations of the replica indices, i.e., $p^\rho=p$, $\hat{p}^\rho=\hat{p}$,
$q^{\rho\delta}=q$, $\hat{q}^{\rho\delta}=\hat{q}$, $\hat{R}^\rho=\hat{R}$ and
$\hat{Q}^{\rho}=\hat{Q}$. With this prescription the evaluation of Eq. (\ref{f_gen})
is straightforward yielding the following replica-symmetric free energy density 
\begin{eqnarray}\label{f1}
-\beta f &=& - \frac{\beta q y}{2} - \beta Q_0 \hat{R} + a + a
\ln(\frac{Q}{a}) + \frac{1-a}{2} \ln \left ( \frac{\pi}{2 \beta (2u-y)}\right ) \nonumber\\
&& + \beta (1-a) \frac{ \hat{R}^2 + q}{2(2u-y)}
+ (1-a) \int_{-\infty}^{\infty} Dz \ln  
\mbox{erfc} \left[ {\frac{\sqrt{\beta} (\hat{R}+z\sqrt{q})}{\sqrt{2(2u-y)}}}
\right] 
\end{eqnarray} 
where $y= \beta (p-q)$ and $Dz=dz \exp(-z^2/2)/\sqrt{2\pi}$ is the Gaussian measure. 
Already at this stage we can see that the concentration of species eliminated
at the outset, given by the parameter $Q$, decouples from the other 
physical parameters and hence does not have any  effect upon them.
In the zero-temperature limit the
saddle-point equations $\partial f/\partial q=0$, $\partial f/\partial y=0$
and $\partial f/\partial \hat{R}=0$ are given by
\begin{equation}\label{sp1}
\Delta = 2 \frac{\sqrt{q}}{Q_0}(u-y) ,
\end{equation}
\begin{equation}\label{sp2}
2y(2u-y)=(1-a) \; \mbox{erfc}\left( -\Delta/\sqrt{2} \right),
\end{equation}
and
\begin{equation}\label{sp3}
\frac{\left (1-a \right )\Delta}{\sqrt{2\pi}}\exp \left( -\Delta^2 /2 \right)= 
\left ( 2u-y \right)^2  - \left ( 2u-y \right ) \left( \Delta^2+1 \right ) y.
\end{equation}
We note that the parameter associated to the concentration of surviving
species $Q_0$ appears only as a scale of $q$ and so henceforth we will set
$Q_0 = 1$ without loss of generality.
In the replica-symmetric framework the Edwards-Anderson order parameter $q$ is defined 
by 
\begin{equation}\label{q_mean}
q = \left \langle \frac{1}{N} \sum_i \langle x_i \rangle_T^2 
\right \rangle .
\end{equation}
If the concentrations $x_i$ were normalized to $1$ rather than to $N$
then $q$ would give the probability that two randomly selected individuals are
of the same species, a quantity known as  Simpson's index \cite{Simp}. 
Nevertheless,
we can still give a simple physical interpretation to $q$. For instance,
values of $q$ of order of $1$ indicate the coexistence of a 
macroscopic number of species (i.e., $x_i \approx 1$ for an extensive
number of species), while large values of $q$ signalize the
dominance of a few species only (i.e., $x_i \approx N$ for a finite
number of species). Of course, this interpretation is equivalent
to that given above for Simpson's index, and so we can view $1/q$
as a measure of the diversity of the ecosystem.
In Fig.\ \ref{fig:q} we present $1/q$  as a function of the dilution
parameter $a$ for several values of the cooperation  pressure $u$. 
The results
of the  numerical  solution of the  replicator equation,
Eq.\  (\ref{eq:rep}), for $N=500$ are also presented. Each data point is
the average over $100$ realizations of the matrix of coupling
strengths, starting with an uniform distribution of concentrations. 
Since the labeling of the species is arbitrary we
can set $b_i =0$ for  $i \leq aN$ and $b_i = 1$ otherwise, without loss
of generality. In addition,  we choose $a$ such that $aN$ is integer
for simplicity.
In agreement with our interpretation, for $a$ close to $1$ 
we can observe the vanishing of $1/q$, which characterizes an
ecosystem composed of a few species only.  For small values
of $u$   the analytical  results show the existence of a maximum of 
diversity for a nonzero value of the dilution parameter (see
the inset in Fig.\  \ref{fig:q}); the numerical
results however do not corroborate this finding. This discrepancy can be explained
by the instability of the replica-symmetric solution. In fact, carrying
out the standard local stability analysis \cite{AT},  we find that this
solution is locally stable wherever the condition 
\begin{equation}
\lambda=-1+\frac{1}{2(2u-y)^2}\mbox{erfc}\left( -\frac{\Delta}{\sqrt{2}}
\right) < 0 
\end{equation}
is satisfied. Figure \ref{fig:AT} shows the regions in the plane
$(a,u)$ where the replica-symmetric solution is  stable. In particular,
we find that for $a=0$  this solution is stable for $u > 1/\sqrt{2}$ while
for $a=1$ it is stable for $u > 1/2$. Hence, the maxima observed in 
Fig.\  \ref{fig:q} are indeed artifacts of the replica-symmetry framework.
Nevertheless, the agreement between the analytical and numerical results
is already excelent  for $u > 0.6$. 
The rather puzzling independence of the diversity on the dilution parameter
for small $u$  has a simple explanation as will be seen in the next
section.

%

\section{Discussion} \label{sec:level4}

\begin{figure}
\includegraphics{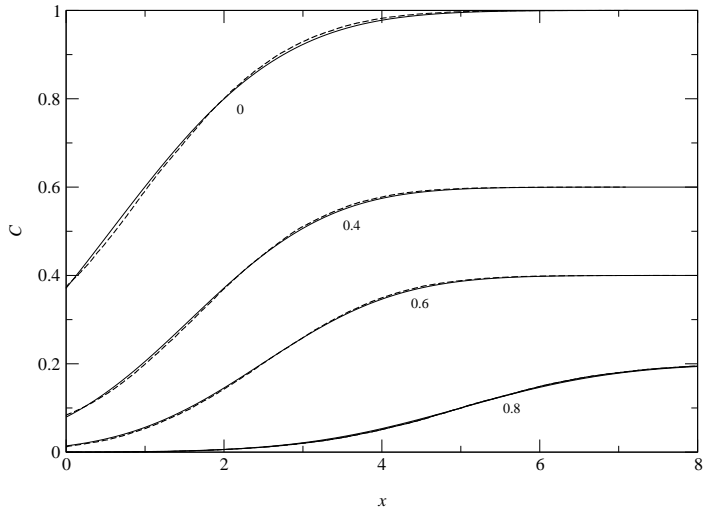} 
\vspace{6cm}
\caption{Cumulative distribution of the concentration of the initially
surviving species in equilibrium for $u=0.8$ and (top to bottom)
$a=0$, $0.4$, $0.6$, 
and $0.8$. The dashed curves are the results of the numerical solution
of the replicator equation.} 
\label{fig:Cx_0.8}
\end{figure}

\begin{figure}
\vspace{1cm}
\includegraphics{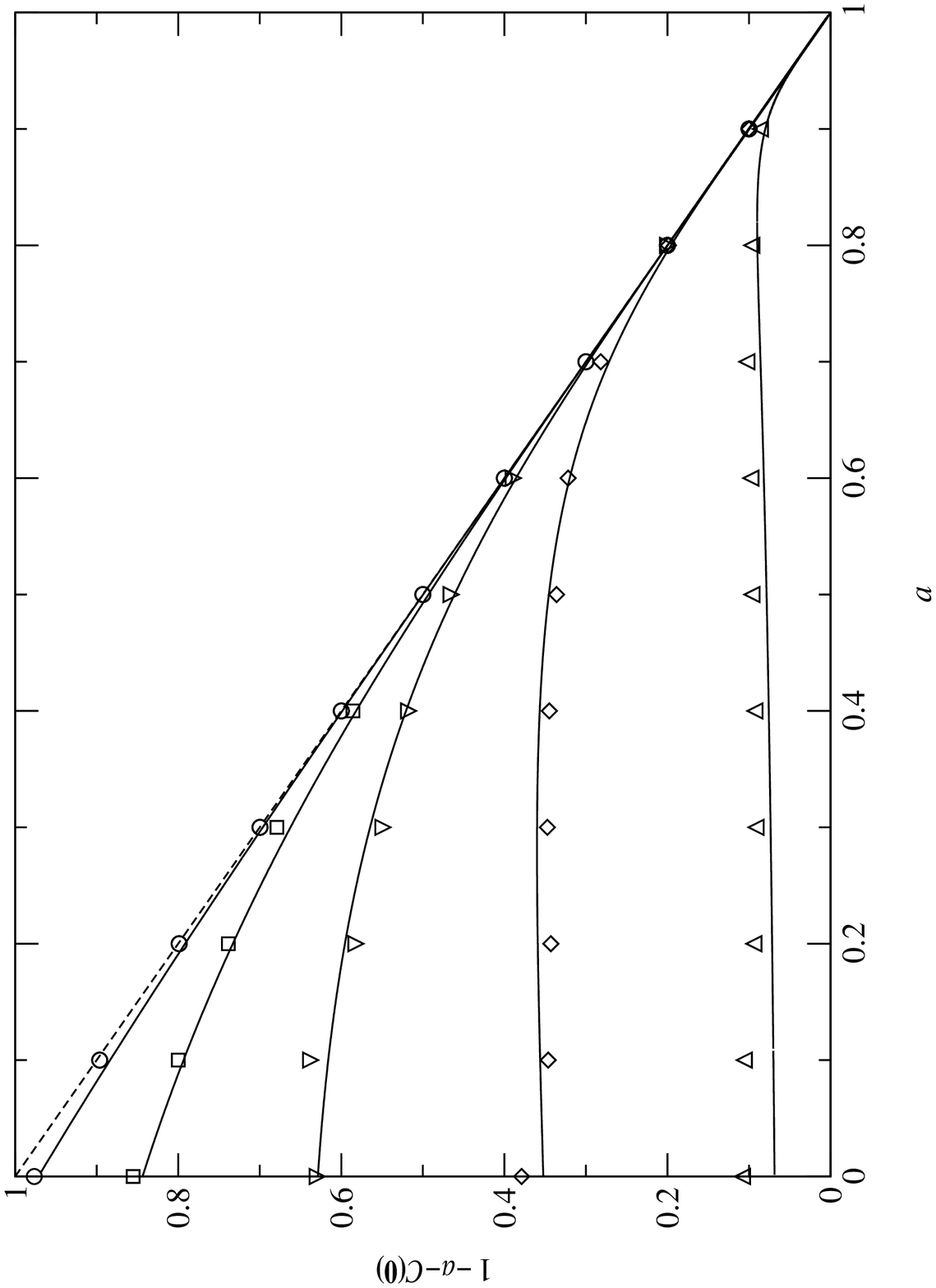} 
\vspace{7cm}
\caption{Fraction of surviving species as a function of the dilution parameter 
$a$ for (top to bottom) $u=1.3 ~(\bigcirc), 1.0 ~(\Box), 0.8 ~(\bigtriangledown),  0.6
~(\Diamond)$ and  $0.3 ~(\bigtriangleup)$. The symbols
are the results of the numerical solution of the
replicator equation. The dashed line is the fraction of species at the
beginning of the coevolutionary process.}
\label{fig:C0}
\end{figure}

Although the interpretation of the reciprocal of the 
Edwards-Anderson order parameter as  
the ecosystem diversity yields some
information on the distribution of species at equilibrium, 
a better understanding is achieved by calculating explicitly
the probability distribution that the concentration of one of the $(1-a)N$ 
remaining species, say $x_k$, assumes the value $x$, defined by 
\begin{equation}\label{prob_x}
{\mathcal{P} }_k \left ( x \right )
 =    \lim_{\beta \rightarrow \infty} \left \langle ~
  \int_0^\infty \prod_j dx_j ~
b_k \delta \left ( x_k - x \right )
{\mathcal W} \left ( {\mathbf x } \right )  \right \rangle
\end{equation} 
with ${\mathcal W} \left ( {\mathbf x } \right )$ given by 
Eq.\ (\ref{Gibbs}). As all non-vanishing species concentrations are equivalent we
can write ${\mathcal{P} }_k \left ( x \right ) = 
{\mathcal{P} }\left ( x \right ) \forall k$. 
Hence to evaluate Eq.\ (\ref{prob_x}) we introduce the auxiliary energy 
\begin{equation}
{\cal H}_{aux} \left ( {\bf x} \right )  =  {\cal H} \left ( {\bf x} \right )
+ h \sum_k b_k \delta \left ( x_k - x \right )  ,
\end{equation}
so that
\begin{equation}\label{der}
{\mathcal{P} }\left ( x \right ) = - \lim_{\beta \rightarrow \infty}  ~ 
\frac{1}{N \beta} \left. \frac{ \partial \langle \ln
Z_{aux} \rangle }{\partial h} \right |_{h=0}
\end{equation}
where $Z_{aux}$ is the partition function (\ref{Z0}) with ${\cal H}$ 
replaced by ${\cal H}_{aux}$.  Using Eq.\ (\ref{der}) the 
calculations needed to evaluate
${\mathcal P} \left ( x \right )$ become analogous to those used 
in the evaluation of the free-energy density (\ref{f1}). In addition,
to handle a possible singularity in the limit 
$\beta \rightarrow \infty$ it is more convenient to deal with
the cumulative distribution function 
${\mathcal{C} }\left ( x \right )  = 
\int_0^{x} dx' \, {\mathcal{P} }\left ( x' \right )$.
Carrying out the calculations within the 
replica-symmetric framework we obtain
\begin{equation}\label{cum}
{\mathcal{C}}(x)= (1-a)\left\{ 1- \frac{1}{2} \mbox{erfc} \left[
\frac{1}{\sqrt{2}} \left( \frac{x(2u-y)}{\sqrt{q}}- \Delta \right) \right]
\right\}
\end{equation}
where $q$, $y$ and $\Delta$ are given by the saddle-point equations
(\ref{sp1})-(\ref{sp3}).
In Fig.\ \ref{fig:Cx_0.8} we show ${\mathcal{C}}(x)$ for 
$u=0.8$  and several values of
$a$. The first point to  note is that 
$ \lim_{x \to \infty} {\mathcal{C}}(x) = 1 - a$ yields the fraction of surviving
species at the outset, as expected. In addition, a nonzero value of
${\mathcal{C}}(0)$ indicates  that the probability
distribution ${\mathcal{P}}(x)$ has a delta peak at $x=0$ and so 
${\mathcal{C}}(0)$ actually yields the fraction of the species that survived
the initial externally induced extinction event but that were extinct later
on due to outcompetition.
In the regime of large dilution, say $a > 0.8$ in Fig.\ \ref{fig:Cx_0.8}, 
the cumulative distribution is very small and practically constant for 
small concentrations, indicating that  
no further extinctions have taken place and, furthermore, that
rare species are very unlikely to be found in the ecosystem at equilibrium.
We note that the numerical simulations yield results practically indistinguishable
from the analytical ones.
The rough independence  of the diversity $1/q$ on the dilution parameter $a$
observed in Fig.\ \ref{fig:q} for small $u$ is easily understood with
the aid of the cumulated distributions. In fact, a direct measure of the
ecosystem diversity is given by the fraction of surviving species 
$ 1- a - {\mathcal{C}}(0)$, which is shown in Fig.\ \ref{fig:C0} as function of $a$.
(We recall that $a$ is the fraction of  species that were extinct 
at the outset due to some external stress and ${\mathcal{C}}(0)$ is the fraction that
died out due to outcompetition.)
The remarkable similarity between these figures  corroborates
our interpretation of $1/q$ as a measure of the diversity. 
Clearly, the diversity is insensitive to variations of $a$
whenever the fraction of extinct species in the undisturbed
ecosystem (i.e. ${\mathcal{C}}(0)$ calculated at $a=0$) is
already considerably  larger than $a$, so that the species
eliminated at the outset would probably be extinct later on 
anyway.


%
\section{Conclusion}\label{sec:level5} 
%

Although the dynamics of the random replicator model may not look
very appealing, in the sense that it always leads  to fixed
points, the frustration caused by the competition between the
concentration limiting self-interactions ($u>0$) and the tendency
to unlimited growth of pairs of strongly cooperative ($J_{ij} < 0$) species 
results in a highly nontrivial equilibrium, characterized by
many meta-stable states \cite{Opper} and a phase of replica symmetry 
breaking \cite{Biscari}. Of course, these very features make some 
aspects of the dynamics (e.g., slow relaxation and hysteresis
effects) nontrivial as well. The wealth of ecologically 
relevant issues that can be addressed within this equilibrium framework
can be appreciated, for instance, in the case
of high-order interactions among the species where it has 
been reported the emergence of a 
threshold  value which gives a lower bound to the concentration of 
the surviving species, preventing then  the existence of rare (low concentration) 
species in the ecosystem \cite{Oliveira}.

An important outcome of the equilibrium analysis 
of  the random replicator model
is the finding that in order to reduce the degree of frustration a fraction of 
the species dies out \cite{Opper}. This type of extinction has clearly
a biotic cause, namely, outcompetition \cite{Maynard}. In this paper
we study how the model ecosystem copes with  abiotic or 
externally induced extinction, in which a fraction of 
randomly chosen species is eliminated at the beginning of
the coevolutionary process. We find that in the regime of high competition
(small $u$) the ecosystem diversity, i.e., the fraction of surviving species is
practically insensitive to the strength $a$ of the initial extinction provided
it is not too large, while in the less competitive regime  (large $u$) the
diversity decreases linearly with increasing $a$. In the case of a large extinction
event we find that no further (biotic) extinctions  take place and, furthermore, 
that rare 
species become very unlikely  to be found in the ecosystem at equilibrium. 
This is distinct from the result mentioned above for the case of
high-order interactions where 
the probability of finding rare species in the ecosystem is 
strictly null \cite{Oliveira}.

An interesting by-product of our investigation is the finding that the 
reciprocal of the Edwards-Anderson order parameter (i.e., the
replica-symmetric overlap between two equilibrium states)  
serves as an easy-to-calculate  measure of the diversity of the 
model ecosystem. This opens the exciting possibility of 
interpreting the different hierarquical levels of the
overlap order parameter in the full replica symmetry breaking
scheme \cite{MPV} as different levels of a phylogenetic tree
that gives the relations of dependence (viewed as ancestrality) among the 
species.

\vspace{1cm}
\noindent
{\large\bf{Acknowledgments}}
\vspace{0.5cm}

The work of J.F.F. is supported in part by Conselho
Nacional de Desenvolvi mento Cient\'{\i}fico e Tecnol\'ogico (CNPq)
and  Funda\c{c}\~ao de Amparo \`a Pesquisa do Estado de S\~ao Paulo 
(FAPESP), Proj.\ No.\ 99/09644-9. V.M.O. is supported by FAPESP.




\end{document}